\documentclass[journal]{IEEEtran}
\usepackage{makecell}
\usepackage{xcolor,colortbl}
\usepackage{color}
\usepackage{colortbl}

\usepackage{multirow,bigdelim}
\usepackage{bm}
\usepackage{subfigure}
\usepackage{nccmath}
\usepackage{float}

\usepackage{amsmath}
\usepackage{hyperref}
\input{def.set}
\usepackage{caption} 
\usepackage{booktabs}
\usepackage{bbding}
\usepackage{pifont}
\usepackage{wasysym}
\usepackage{amssymb}
\usepackage{graphicx}
\usepackage{booktabs,amsfonts,dcolumn}

\usepackage{algorithmic, algorithm}
\usepackage{mathtools}
\usepackage{tipa}

\newcommand{\zhenk}[1]{{\color{black}{#1}}}
\newcommand{\zhenksep}[1]{{{#1}}}
\newcommand{\xmark}{\ding{55}}%
\newcommand{\tbd}{\ding{108}}%

\usepackage{scalerel,stackengine}
\usepackage{algorithmic, algorithm}

\stackMath
\newcommand\reallywidehat[1]{%
\savestack{\tmpbox}{\stretchto{%
  \scaleto{%
    \scalerel*[\widthof{\ensuremath{#1}}]{\kern-.6pt\bigwedge\kern-.6pt}%
    {\rule[-\textheight/2]{1ex}{\textheight}}
  }{\textheight}%
}{0.5ex}}%
\stackon[1pt]{#1}{\tmpbox}%
}
\ifCLASSINFOpdf
\else
\fi
\begin{document}
%
\title{Cascaded Cross-Module Residual Learning towards Lightweight End-to-End Speech Coding}

\title{Cross Module Residual Learning And Collaborative Quantization for Efficient Neural Waveform Coding}

\title{Scalable and Efficient Neural Waveform Coding with Cascaded Coding and Collaborative Quantization}

\title{Neural Waveform Coding with Cascaded Coding And Collaborative Quantization}
\title{Revitalizing Multistage Quantization And Linear Prediction For Efficient Neural Waveform Coding}

\title{Cascaded Coding With Collaborative Quantization for Efficient Neural Waveform Compression }

\title{Efficient Speech Waveform Coding With Residual Cascading And Collaboratively Quantized LPC Preprocessing}

\title{Collaboratively Quantized LPC With Residual Cascading for Efficient Speech Waveform Coding}


\title{Deep Learning in Speech Waveform Coding}

\title{Residual Cascading With Collaborative Quantization Towards Efficient Blockwise  Neural Speech Coding}
\title{Scalable Neural Waveform Coding for Speech Compression: A Lightweight Design}
\title{Neural Waveform Coding: A Lightweight Design}
\title{DSP Adapted Neural Waveform Coding}
\title{Scalable and Efficient Neural Speech Coding}
\title{Scalable and Efficient Neural Speech Coding: \\A Hybrid Design}

%
%
%

\author{Kai~Zhen,~\IEEEmembership{Student~Member,~IEEE,}~
        Jongmo Sung,~
        Mi Suk Lee,~
        Seungkwon Beack,\\
        Minje~Kim,~\IEEEmembership{Senior~Member,~IEEE,}
\thanks{This work was supported by the Institute for Information and Communications Technology Promotion (IITP) funded by the Korea government (MSIT) under Grant 2017-0-00072 (Development of Audio/Video Coding and Light Field Media Fundamental Technologies for Ultra Realistic Tera-Media).
Kai Zhen is with the Department of Computer Science and Cognitive Science Program at Indiana University, Bloomington, IN 47408 USA. Jongmo Sung, Mi Suk Lee, and Seungkwon Beack are with Electronics and Telecommunications Research Institute, Daejeon, Korea 34129. Minje Kim is with the Dept. of Intelligent Systems Engineering at Indiana University (e- mails: zhenk@iu.edu, lms@etri.re.kr, jmseong@etri.re.kr, skbeack@etri.re.kr, minje@indiana.edu).}
\thanks{Manuscript received March XX, 2021; revised YYY ZZZ, 2021.}}

%
%

\markboth{IEEE/ACM Transactions on Audio, Speech, and Language Processing,~Vol.~XX, No.~YY, September~2021}%
{Zhen \MakeLowercase{\textit{et al.}}: Scalable and Efficient Neural Speech Coding}
%



\maketitle

\begin{abstract}
We present a scalable and efficient neural waveform coding system for speech compression. We formulate the speech coding problem as an autoencoding task, where a convolutional neural network (CNN) performs encoding and decoding as a neural waveform codec (NWC) during its feedforward routine. The proposed NWC also defines quantization and entropy coding as a trainable module, so the coding artifacts and bitrate control are handled during the optimization process. We achieve efficiency by introducing compact model components to NWC, such as gated residual networks and depthwise separable convolution. Furthermore, the proposed models are with a scalable architecture, cross-module residual learning (CMRL), to cover a wide range of bitrates. To this end, we employ the residual coding concept to concatenate multiple NWC autoencoding modules, where each NWC module performs residual coding to restore any reconstruction loss that its preceding modules have created. CMRL can scale down to cover lower bitrates as well, for which it employs linear predictive coding (LPC) module as its first autoencoder. The hybrid design integrates LPC and NWC by redefining LPC’s quantization as a differentiable process, making the system training an end-to-end manner. The decoder of proposed system is with either one NWC (0.12 million parameters)  in low to medium bitrate ranges (12 to 20 kbps) or two NWCs  in the high bitrate (32 kbps). Although the decoding complexity is not yet as low as that of conventional speech codecs, it is significantly reduced from that of other neural speech coders, such as a WaveNet-based vocoder. For wide-band speech coding quality, our system yields comparable or superior performance to AMR-WB and Opus on TIMIT test utterances at low and medium bitrates. The proposed system can scale up to higher bitrates to achieve near transparent performance.
\end{abstract}





\section{Introduction}
\IEEEPARstart{S}{peech} coding can be implemented as an encoder-decoder system, whose goal is to compress input speech signals into the compact bitstream (encoder) and then to reconstruct the original speech from the code with the least possible quality degradation.  Speech coding facilitates telecommunication and saves data storage among many other applications. There is a typical trade-off a speech codec must handle: the more the system reduces the amount of bits per second (bitrate), the worse the perceptual similarity between the original and recovered signals is likely to be perceived. In addition, the speech coding systems are often required to maintain an affordable computational complexity when the hardware resource is at a premium. 

For decades, speech coding has been intensively studied yielding  various standardized codecs that can be categorized into two types: the vocoders and waveform codecs. A vocoder, also referred to as parametric speech coding, distills a set of physiologically salient features, such as the spectral envelope (equivalent to vocal tract responses including the contribution from mouth shape, tongue position and nasal cavity), fundamental frequencies, and gain (voicing level), from which the decoder \textit{synthesizes} the speech. Typically, a vocoder operates at $3$ kbps or below with high computational efficiency, but the synthesized speech quality is usually limited and does not scale up to higher bitrates \cite{valin2019real}\cite{schroeder1966vocoders}\cite{mccree1995mixed}. On the other hand, a waveform codec aims to accurately reconstruct the input speech signal, which features up-to-transparent quality in a high bitrate range \cite{dietz2015overview}. 
\zhenksep{AMR-WB \cite{amrwb}, for instance, can be seen as a hybrid waveform codec, because it employs speech modeling as in many other waveform codecs \cite{stachurski2000combining}\cite{schroeder1985code}\cite{stachurski20004}. EVS \cite{bruhn2015standardization}, a recently standardized 3GPP voice and audio codec, has noticeably optimized frame error robustness, yielding a much-enhanced frame error concealment performance against than AMR-WB \cite{ramo2015subjective}. Similar to EVS, Opus, a waveform codec at its core, can also be applied to both speech and audio signals where it uses the LPC-based SILK algorithm for the speech-oriented model \cite{skoglund2019improving} and scales up to 510 kbps for transparent audio streaming and archiving.}


Under the notion of unsupervised speech representation learning, deep neural network (DNN)-based codecs have revitalized the speech coding problem and provided different perspectives \cite{OordA2017vqvae}\cite{chorowski2019unsupervised}. 
The major motivation of employing neural networks to speech coding is twofold: to fill the performance gap between vocoders and waveform codecs towards a near-transparent speech synthesis quality; to use its trainable encoder and learn latent representations which 
may benefit other DNN-implemented downstream applications, such as speech enhancement \cite{xu2014regression}\cite{luo2018tasnet}, speaker identification \cite{nagrani2017voxceleb} and automatic speech recognition \cite{graves2013speech}\cite{abdel2014convolutional}. Having that, a neural codec can serve as a trainable acoustic unit integrated in future digital signal processing engines \cite{chorowski2019unsupervised}.




Recently proposed neural speech codecs have achieved high coding gain and reasonable quality by employing deep autoregressive models. 
The superior speech synthesis performance achieved in WaveNet-based models \cite{van2016wavenet} has successfully transferred to neural speech coding systems, such as in \cite{KleijnW2018wavenet}, where WaveNet serves as a decoder synthesizing wideband speech samples from a conventional non-trainable encoder at  $2.4$ kbps. Although its reconstruction quality is comparable to waveform codecs at higher bitrates, the computational cost is significant due to the model size of over $20$ million parameters. 

Meanwhile, VQ-VAE \cite{OordA2017vqvae} integrates a trainable vector quantization scheme into the variational autoencoder (VAE) \cite{kingma2013auto} for discrete speech representation learning. While the bitrate can be lowered by reducing the sampling rate  $64$ times, the downside for VQ-VAE is that the prosody can be significantly altered. Although \cite{GarbaceaC2019vqvae} provides a scheme to pass the pitch and timing information to the decoder as a remedy, it does not generalize to non-speech signals. More importantly, VQ-VAE as a vocoder does not address the complexity issue since it uses WaveNet as the decoder. Although these neural speech synthesis systems noticeably improve the speech quality at low bitrates, they are not feasible for real-time speech coding on the hardware with limited memory and bandwidth. 


LPCNet \cite{valin2019lpcnet} focuses on efficient neural speech coding via a WaveRNN \cite{wavernn} decoder by leveraging the traditional linear predictive coding (LPC) techniques. The input of the LPCNet is formed by $20$ parameters ($18$ Bark scaled cepstral coefficients and $2$ additional parameters for the pitch information) for every $10$ millisecond frame. All these parameters are extracted from the non-trainable encoder, and vector-quantized with a fixed codebook.
As discussed previously, since LPCNet functions as a vocoder, the decoded speech quality is not considered transparent \cite{valin2019real}. 

\begin{table}[t]
\setlength\tabcolsep{.4pt}
\caption{Categorical summary of recently proposed neural speech coding systems. \CheckmarkBold~means the system supports the feature while \xmark~ does not. \tbd~means it is not known.}
\centering
\resizebox{.999\columnwidth}{!}{%
\begin{tabular}{c||cccc}
  \toprule
  & WaveNet \cite{KleijnW2018wavenet} & VQ-VAE \cite{GarbaceaC2019vqvae} & LPCNet \cite{valin2019lpcnet} & Proposed \\\midrule
Transparent coding &\CheckmarkBold  &\tbd  &\xmark  &\CheckmarkBold  \\
Less than 1M parameters &\xmark  &\xmark  &\CheckmarkBold  &\CheckmarkBold \\
Real-time processing  &\xmark  &\xmark  &\CheckmarkBold  &\CheckmarkBold \\
Encoder trainable &\CheckmarkBold  &\CheckmarkBold  &\xmark  &\CheckmarkBold  \\
\bottomrule
\end{tabular}}
\label{tab:codec_comp}
\end{table}

\zhenk{In this paper, we propose a novel neural speech coding system, with a lightweight design and scalable performance. First, we design a generic neural waveform codec with only $0.35$ million parameters where $0.12$ million parameters belong to the decoder. Compared to our previous models in \cite{zhen2019cascaded}\cite{zhen2020efficient} where the decoder has $0.23$ million parameters, the current neural codec employs gated linear units to boost the gradient flow during model training and depthwise separable convolution to achieve further efficiency during decoding, as detailed in Sec. \ref{sec:neuralcodec}.}
Based on this neural codec, our full system features two mechanisms to integrate speech production theory and residual coding techniques in Sec. \ref{sec:dsp}. Benefited from the residual-excited linear prediction (RELP) \cite{un1975residual},  we conduct LPC and apply the neural waveform codec to the excitation signal, which is illustrated in Sec.\ref{sec:cq}. In this integration, a trainable quantizer bridges the encoding of linear spectral pairs and the corresponding LPC residual, making the speech coding pipeline end-to-end trainable. 
We also enable residual coding among neural waveform codecs to scale up the performance for high bitrates (Sec.\ref{sec:cmrl}). 
\zhenk{In summary, the proposed system has following characteristics:}
\begin{itemize}
    \item \textit{Scalability}: Similar to LPCNet \cite{valin2019lpcnet}, the proposed system is compatible with conventional spectral envelope estimation techniques. However, ours operates at a much wider bitrate range with comparable or superior speech quality to standardized waveform codecs. 
    \item \textit{Compactness}: The neural waveform codec in our system is with a much lower complexity than WaveNet \cite{van2016wavenet} and VQ-VAE \cite{OordA2017vqvae} based codecs. Our decoder contains only $0.12$ million parameters which is $160\times$ more compact than a WaveNet counterpart. \zhenk{Our TensorFlow implementation's execution time to encode and decode a signal is only $42.44\%$ of its duration on a single-core CPU in the low-to-medium bitrates and 80.21\% in the high bitrate}, facilitating real-time communications.
    \item \textit{Trainability}: Our method is with a trainable encoder as in VQ-VAE, which can be integrated into other DNNs for acoustic signal processing. Besides, it is not constrained to speech, and can be generalized to audio coding with minimal effort as shown in \cite{zhenk2020nac}. 
\end{itemize}
TABLE \ref{tab:codec_comp} highlights the comparison to the other existing neural speech codecs.

\zhenk{This paper is an extension of the authors' previous conference presentations \cite{zhen2019cascaded}\cite{zhen2020efficient}, where some initial ideas were already discussed. The new contributions presented this journal version are listed as follows:
\begin{itemize}
    \item \textit{Novel algorithmic enhancements}: We propose new neural network architectures to form a new baseline autoencoder module and used it everywhere in our codecs. In our previous works, we have used a 1D convolutional neural network (CNN) that defines an autoencoder block with an identity shortcut as in the ResNet architecture \cite{he2016deep}. While this architecture has been effective, in this journal paper, we propose to use the dilated gated linear units and depthwise separable convolution to reduce the kernel size without inducing any performance degradation. Consequently, our NWC is defined by 0.35M parameters, whose decoder part accounts for only 0.12M parameters. Compared to our previous models that are already small with only 0.45M parameters, the newly introduced reduction amounts to $22.2\%$. If we only compare the decoder parts, it is a $47.8\%$ reduction. Although the proposed architecture is more compact than our previous models or the WaveNet-based codecs, since neural codecs' complexity is much larger than the traditional speech codecs, the additional model complexity reduction with no degradation of performance is promising. The architectural improvement are presented in Sec. \ref{sec:nwc_arc}.
    \item \textit{Extensive experimental validation}: In our previous works, the experimental validation was to prove the initial concepts individually proposed each paper. In this time, we conduct an extensive and thorough experiments to provide the readers with a full view to the whole building-blocks of the neural speech coding. To this end, we define four candidate systems, from Model-I to IV, by incrementally adding new modules, such as LPC, the trainable LPC quantizer, and multiple concatenated neural autoencoders. The objective and subjective tests validate each of these additions in a full view (TABLE \ref{tab:ob_comp} and Fig. \ref{fig:box}). 
    \item \textit{Additional analyses and ablation tests}: We also provide detailed experimental validation for most of the claims made in the paper by designing and performing separate experiments, which were missing in the previous papers.
    \begin{itemize}
        \item Sec. \ref{sec:nwc_analysis} provides experimental verification that the proposed compact neural architecture does not induce performance loss. 
        \item Sec. \ref{sec:cmrl_analysis} presents a detailed analysis of the behavior of the cascaded autoendoers and the impact of different training strategies. 
        \item Sec. \ref{sec:loss_abl} explores contribution of different loss terms in our training objective by performing ablation tests, and then proposes an optimal combination of hyperparameters. 
        \item Sec. \ref{sec:cq_abl} also conducts an ablation test to empirically verify that the proposed trainable LPC quantization algorithm improves speech quality at the same bitrate. 
        \item Sec. \ref{sec:bit_alloc_lpc_nwc} and \ref{sec:bit_alloc_nwc} analyze the bit allocation behavior among the different submodules. Since the bit allocation strategy is decided by the learning algorithm, these analyses provide evidence that our models dynamically adapt to the characteristics of the signals given the limited bit budget. 
        \item Sec. \ref{sec:complexity} presents additional analyses on computational complexity and execution time ratios to discuss the potential of the neural codecs in real-time applications. 
        \item Last but not least, in Sec. \ref{sec:limitations} we discuss the implementation issues and the limitations of the proposed system in the context of real-world application scenarios.
    \end{itemize}
\end{itemize}
}

 \begin{figure}[t]
\centering
\subfigure[The high-level structure of proposed neural waveform codec]{\includegraphics[width=\columnwidth]{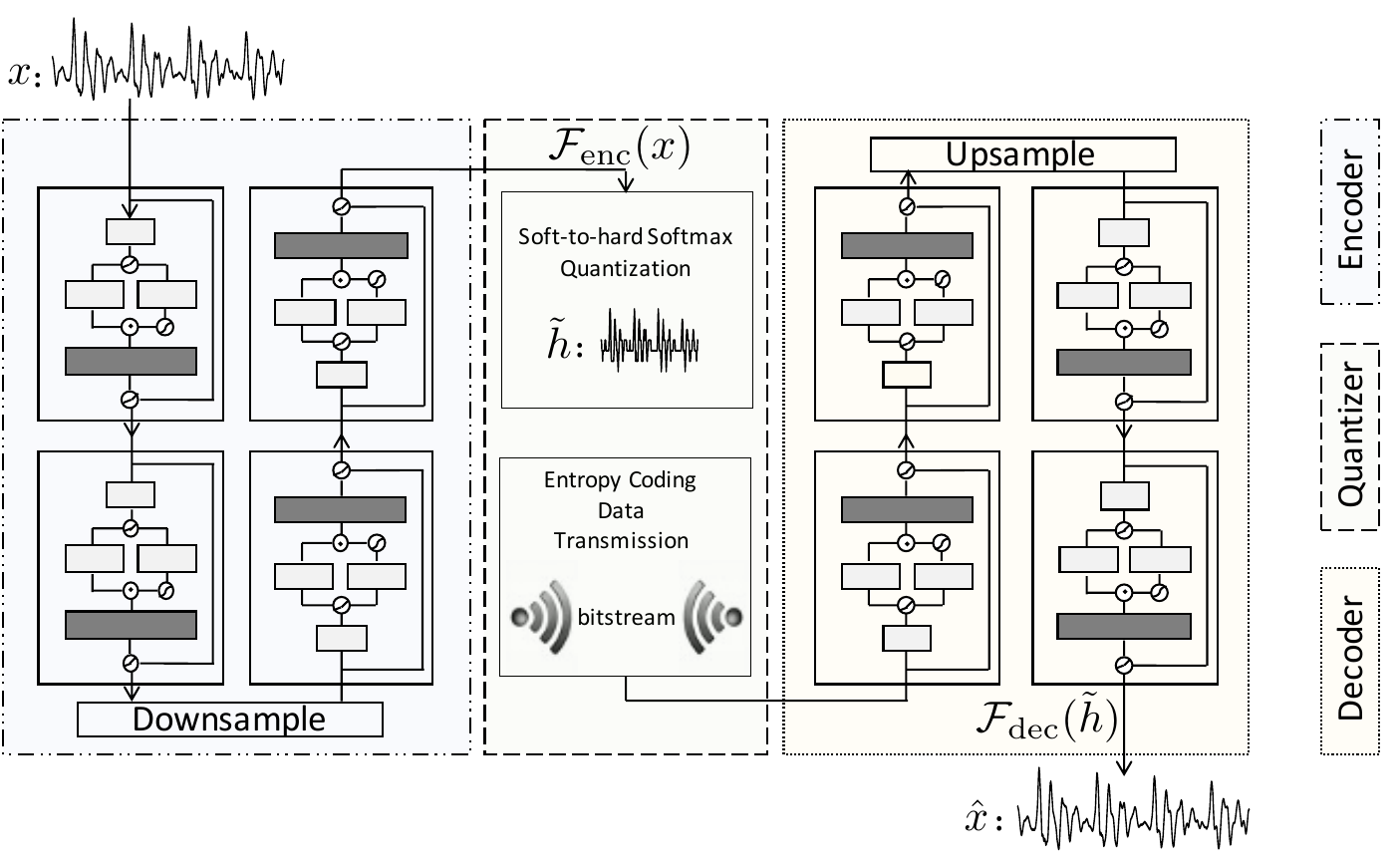}}
\subfigure[Dilated gated linear unit (GLU)]{\includegraphics[scale=0.53]{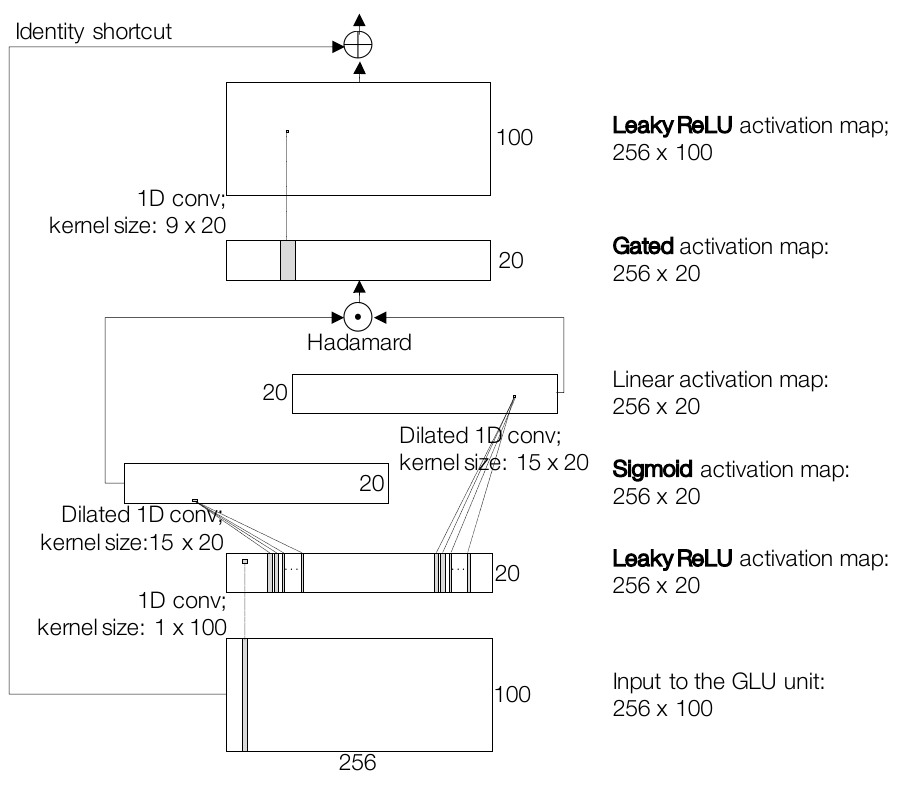}}
\label{fig:dwcnn}
\subfigure[Depthwise separable 1D convolution for upsampling]{\includegraphics[scale=0.53]{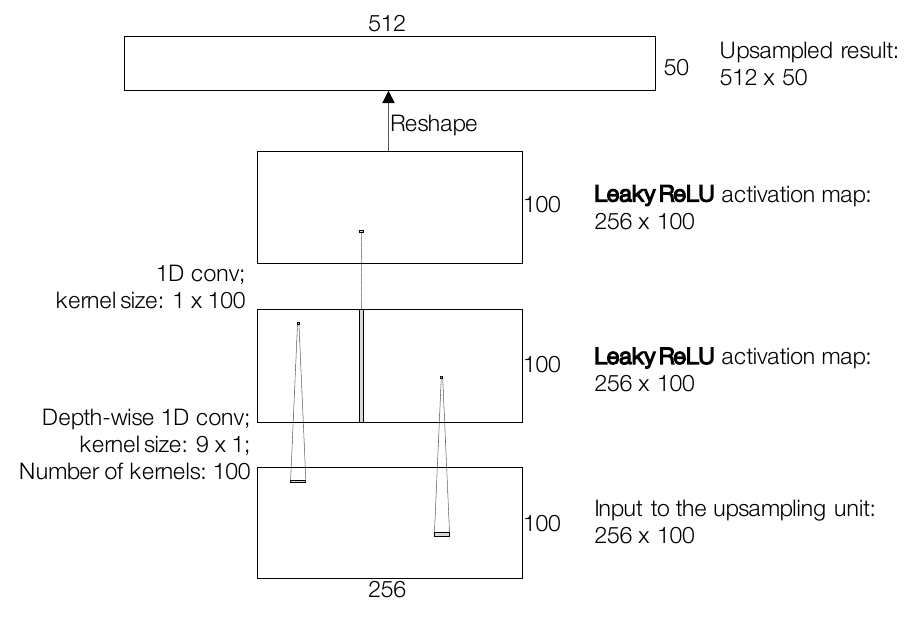}}
\caption{The proposed architecture for lightweight NWC.}
 \label{fig:nc}
 \end{figure}

\section{End-to-End Neural Waveform Codec (NWC)}
\label{sec:neuralcodec}
The neural waveform codec (NWC), is an end-to-end autoencoder that forms the base of our proposed coding systems. 
\zhenk{NWC directly encodes the input waveform $\bx\in\Real^T$ using a convolutional neural network (CNN) encoder $\calF_\text{enc}(\cdot)$, i.e., $ {\bh}\leftarrow\calF_\text{enc}(\bx)$. Then, the quantization process $\calQ(\cdot)$ converts the encoding into a bitstring $\tilde{\bh}\in\Real^N$, which is followed by lossless data compression and bitstream transmission. On the receiver side, the decoder reconstructs the waveform as $\bx\approx\hat{\bx}\leftarrow\calF_\text{dec}(\tilde{\bh})$.
Fig.  \ref{fig:nc} (a) depicts NWC's overall system architecture. The structure is detailed in TABLE \ref{tab:topo}. It serves as a basic component in the proposed speech coding system in Sec. \ref{sec:dsp}. In this section, we first introduce the architectural improvement that reduced our model's complexity. Next, we also introduce two strategies that compress the signals: feature map compression and trainable quantization. }

\zhenk{
\subsection{The Improved Architecture for NWC}
\label{sec:nwc_arc}
We propose two different kinds of structural modification to reduce the model's overall complexity compared to other NWC models including our prior works \cite{zhen2019cascaded,zhen2020efficient}. 

First, both encoder and decoder adopt gated linear units (GLU) \cite{DauphinY2017glu}. We also define the GLU's convolution with dilation \cite{yu2015multi} to expand the receptive field in the time domain, which is a scheme that showed promising performance in speech enhancement \cite{tan2018gated2}. 
Fig. \ref{fig:nc} (b) shows our dilated GLU module. It first reduces the channel from 100 to 20 using a unit-width kernel. Then, two separate dilated convolution layers are applied to produce two feature maps, one of which goes through a sigmoid activation. Hence, the Hadamard product of the two feature maps can be seen as a \textit{gated} version of the linear feature map. It is known that this gating mechanism in the middle boosts the gradient flow thanks to the linear path that does not involve the gradient vanishing issue. The final result is a mixture of the input and the last feature map, turning this block into a residual learning function as proposed in ResNet \cite{he2016deep}. This kind of architecture shows superior performance as evidenced in \cite{DauphinY2017glu, tan2019gated}. 

Our encoder reduces the data rate by using a ``downsampling" convolutional layer, whose ``stride" parameter is set to be 2. As a counterpart, the decoder's ``upsampling" layer makes up the loss. More details about this down and upsampling operations will be discussed in Sec. \ref{sec:downsample}. In this subsection, we focus on the actual module that performs the upmixing, where we introduced additional reduction in complexity. Out of various choices, we employ the depthwise separable convolution \cite{chollet2017xception} to further save the computational cost (Fig.  \ref{fig:nc} (c)). For example, to transform a feature map of size $256\times 100$ (features, channels) into its upsampled version of size  $512\times 50$, we first perform depthwise convolution using a $c\times1$ kernel. In our system $c=9$. Since the depthwise convolution applies to each channel separately, we eventually need $100$ such kernels. It is a reduction of model complexity, because a normal convolution requires a kernel of size $c\times 100\times 100$ (features, input channels, output channels), which is 100 time larger. In depthwise separable convolution, another $1\times1$ convolution follows to add more nonlinearity, for which we need a $1\times 100\times 100$ kernel. In our case, it is easy to show that $c\times 100 \times 100 > c\times 100+100\times 100$ when the integer $c>1$. For example, when $c=9$, it is a reduction of about 88\% of parameters.}

\zhenk{Likewise, the proposed NWC is lightweight with only 0.35 million parameters, which is a reduction of 0.1 million parameters compared to our previous works \cite{zhen2019cascaded}\cite{zhen2020efficient}. The reduction comes from the streamlined upsampling operation implemented via the depthwise separable convolution. Eventually, the decoder accounts for 0.12 million parameters out of 0.35.}

\begin{table}[t]
\centering
\caption{Architecture of the neural waveform codec: input and output tensors are shaped as (sample, channel), while the kernel is represented as (kernel size, in channel, out channel).}
\label{tab:topo}
\setlength\tabcolsep{3.6pt}
\begin{tabular}{ c| c|c|c|c }
 \toprule
 & Layer &Input shape & Kernel shape & Output shape\\
 \midrule
\multirow{13}{*}{\rotatebox[origin=c]{90}{\zhenk{Encoder}}} & \begin{tabular}{@{}c@{}}Channel\\Expansion\end{tabular} & (512, 1) & (55, 1, 100) &(512, 100) \\
\cmidrule{2-5}
& \begin{tabular}{@{}c@{}}Gated\\Linear\\Unit\end{tabular} & (512, 100) & \begin{tabular}{cc}\rule[6pt]{0pt}{0pt}(1, 100, 20)  &\rdelim]{4}{5mm}[$\times$2]\\ (15, 20, 20)$^\dagger$  & \\(15, 20, 20)$^\dagger$  & \\(9, 20, 100) &\rule[-1pt]{0pt}{0pt}\end{tabular} &(512, 100)  \\\cmidrule{2-5}
& Downsampling & (512, 100) & (9, 100, 100) &(256, 100) \\
\cmidrule{2-5}
& \begin{tabular}{@{}c@{}}Gated\\Linear\\Unit\end{tabular} & \zhenk{(256, 100)} & \begin{tabular}{cc}\rule[6pt]{0pt}{0pt}(1, 100, 20)  &\rdelim]{4}{5mm}[$\times$2]\\ (15, 20, 20)$^\dagger$  & \\(15, 20, 20)$^\dagger$  & \\(9, 20, 100) &\rule[-1pt]{0pt}{0pt}\end{tabular} &\zhenk{(256, 100)}  \\\cmidrule{2-5}
& \begin{tabular}{@{}c@{}}Channel\\Reduction\end{tabular} & (256, 100) & (9, 100, 1) &(256, 1) \\
\midrule
\midrule
\multirow{15}{*}{\rotatebox[origin=c]{90}{\zhenk{Decoder}}}& \begin{tabular}{@{}c@{}}Channel\\Expansion\end{tabular} & (256, 1) & (9, 1, 100) &(256, 100) \\
\cmidrule{2-5}
& \begin{tabular}{@{}c@{}}Gated\\Linear\\Unit\end{tabular} & (256, 100) & \begin{tabular}{cc}\rule[6pt]{0pt}{0pt}(1, 100, 20)  &\rdelim]{4}{5mm}[$\times$2]\\ (15, 20, 20)$^\dagger$  & \\(15, 20, 20)$^\dagger$  & \\(9, 20, 100) &\rule[-1pt]{0pt}{0pt}\end{tabular} &(256, 100)  \\\cmidrule{2-5}
& Upsampling & (256, 100) & \begin{tabular}{cc}\rule[6pt]{0pt}{0pt}(9, 100, 1)  &\\ (1, 100, 100) &\rule[-1pt]{0pt}{0pt}\end{tabular} &(512, 50) \\
\cmidrule{2-5}
& \begin{tabular}{@{}c@{}}Gated\\Linear\\Unit\end{tabular} & (512, 50) & \begin{tabular}{cc}\rule[6pt]{0pt}{0pt}(1, 50, 20)  &\rdelim]{4}{5mm}[$\times$2]\\ (15, 20, 20)$^\dagger$  & \\(15, 20, 20)$^\dagger$  & \\(9, 20, 50) &\rule[-1pt]{0pt}{0pt}\end{tabular} &(512, 50)  \\\cmidrule{2-5}
& \begin{tabular}{@{}c@{}}Channel\\Reduction\end{tabular} & (512, 50) & (55, 50, 1) &(512, 1) \\
\bottomrule
\end{tabular}
\vspace{0.15in}
\end{table}

\subsection{Feature Map Compression}\label{sec:downsample}
One way to compress the input signal in the proposed encoder architecture is to reduce the data rate. The CNN encoder function takes an input frame, $\bx\in\Real^{T}$, and converts it into a feature map $\bh\in\Real^{N}$,
\begin{equation}
    \bh\leftarrow\calF_\text{enc}(\bx),
\end{equation} 
which then goes through quantization, transmission, and decoding to recover the input as shown in Fig. \ref{fig:nc} (a). During the encoding process, we introduce a \textit{downsampling} operation, reducing the dimension of the code vector  $\bh$. We employ a dedicated downsampling layer by setting up the stride value to be $2$ during its convolution, reducing the data rate by $50\%$, i.e., $N=T/2$. Accordingly, the decoder needs a corresponding upsampling operation to recover the original sampling rate. We use subpixel CNN layer proposed in \cite{shi2016real} to recover the original sampling rate. 
Concretely, the subpixel upsampling involves a feature transformation implemented in depthwise convolution, and a shuffle operation that interlaces features from two channels into a single channel, as shown in Eq. (\ref{eq:upsampling}), where the input feature of the shuffle operation is shaped as ($N$, $2$) and the output is shaped as ($2N$, $1$).
\begin{equation}
\begin{split}
    &[h_{11}, h_{21}, h_{12}, h_{22}, \ldots, h_{1N}, h_{2N} ]\\
    &\leftarrow \text{Upsampling}([h_{11}, h_{12}, \ldots, h_{1N}; h_{21}, h_{22}, \ldots, h_{2N} ]
\end{split}
\label{eq:upsampling}
\end{equation}  

\subsection{The Trainable Quantizer for Bit Depth Reduction}
The dimension-reduced feature map can be further compressed via bit depth reduction. Hence, the floating-point code $\bh$ goes through quantization and entropy coding, which will finalize the bitrate based on the entropy of the code value distribution.
Typically, a bit depth reduction procedure lowers the average amount of bits to represent each sample. In our case, we could employ a quantization process that assigns the output of the encoder to one of the pre-defined quantization bins. If there are $2^5=32$ quantization bins, for example, a single-precision floating-point value's bit depth reduces from 32 to 5. In addition, various entropy coding techniques, such as Huffman coding, can be further employed to losslessly reduce the bit depth. While the quantization could be done in a traditional way, e.g., using Lloyds-Max quantization \cite{garey1982complexity} \textit{after} the neural codec is fully trained, we encompass the quantization step as a trainable part of the neural network as proposed in \cite{AgustssonE2017softmax}. Consequently, we expect that the codec is aware of the quantization error, which the training procedure tries to reduce it. It is also convenient to control the bitrate by controlling the entropy of the code value distribution, which can be also done as a part of network training.


In NWC, the quantization process is represented as classification on each scalar value of the encoder output. Given a vector with $K$ centroids, $\boldsymbol{\beta}=[\beta_1, \beta_2, \cdots, \beta_K]^\top$, the quantizer's goal is to assign each feature $h_n$ to the closest centroid in terms of $\ell_2$ distance, which is defined as follows:
\begin{equation}
\label{eq:dist}
    \bD =\begin{bmatrix} 
    ||h_1-\beta_1||_2 & \cdots & ||h_1-\beta_K||_2\\
    \vdots & \ddots  &\vdots\\
    ||h_N-\beta_1||_2 &   \cdots     & ||h_N-\beta_K||_2
    \end{bmatrix},
\end{equation}
where $n$-th row in $\bD$ is a vector of $\ell_2$ distance between $n$-th code value $h_n$ to all $K$ quantization bins. Then, we employ the softmax function to turn each row of $\bD$ into a $K$-dimensional probabilistic assignment vector: 
\begin{equation}
\label{eq:soft}
    \bA^{\text{(soft)}} =\begin{bmatrix} 
    \text{softmax}(-\alpha\bD_{1:})\\
    \text{softmax}(-\alpha\bD_{2:})\\
    \vdots\\
    \text{softmax}(-\alpha\bD_{N:})\\
    \end{bmatrix},
\end{equation}
where we turn the distance into a similarity metric by multiplying a negative number $-\alpha$, such that the shortest distance is converted to the largest probability. \zhenk{In our implementation, $\boldsymbol{\beta}$ is initialized as a vector of $K=32$ uniformly spaced numbers within the interval of $[-1, 1]$. As for $\alpha$, we begin with a large enough number $300$. Both $\alpha$ and $\beta$ are trainable parameters to optimize the quantization process.}

Note that Eq. \eqref{eq:soft} yields a soft assignment matrix $\bA^{\text{(soft)}} \in \Real^{N\times K}$. In practice, though, the quantization process must perform a hard assignment, so each code value $h_n$ is replaced by an integer index to the closest centroids:  $z_n\in\{1,2,\cdots,K\}$, which is represented by $\lceil\log_2 K\rceil$ bits as the quantization result. The hard kernel assignment matrix  $\bm{A}^{\text{(hard)}}$, where each row is a one-hot vector, can be induced by turning on the maximum element of $\bA^{\text{(soft)}}$ while suppressing the non-maximum:
\begin{equation}
    \bA^{\text{(hard)}}_{nk}=\left\{\begin{array}{rl}
    1 & \text{ if } \argmax_{j\in\{1,2,\cdots,K\}} \bA^{\text{(soft)}}_{nj} = k\\
    0 & \text{otherwise}
    \end{array}\right.
    \label{eq:hard3}
\end{equation}
On the decoder side, $\tilde{\bh}=\bA^{\text{(hard)}}\bbeta$ recovers $\bh$.



\begin{algorithm}[t]
\caption{\zhenk{Soft-to-hard quantization during inference}, $\mathcal{Q}(\bh, \alpha, \boldsymbol{\beta})$}
\label{algo:quan}
\begin{algorithmic}[1]
\STATE \textbf{Input:} the code, e.g., the encoder output, $\bh=\mathcal{F}_{\text{enc}}(\bx)$\\
\hspace{0.38in} the \zhenk{trained} softmax scaling factor, $\alpha$\\
\hspace{0.38in} the \zhenk{trained} centroid vector, $\boldsymbol{\beta}\in\Real^{K}$
\STATE \textbf{Output:} the quantized code, $\hat{\bh}$ (training) or $\tilde{\bh}$ (testing)
\STATE Compute the dissimilarity matrix: $\bD_{nk}\leftarrow \ell_2(h_n||\beta_k)$
\STATE Softmax conversion:  $\bA^{\text{(soft)}}_{n:} \leftarrow \text{Softmax}(-\alpha\bD_{n:})$
\IF {Training} 
    \STATE Soft quantization:  $\hat{\bh} \leftarrow \bA^{\text{(soft)}}\boldsymbol{\beta}$
\ELSIF {Testing}
    \STATE Hard quantization:  $\tilde{\bh} \leftarrow \bA^{\text{(hard)}}\boldsymbol{\beta}$
\ENDIF 
 \end{algorithmic}
\end{algorithm}

Since $\arg\max$ operation in Eq. \eqref{eq:hard3} is not differentiable, a soft-to-hard scheme is proposed in \cite{AgustssonE2017softmax}, where $\bA^{\text{(hard)}}$ is used only at test time. During backpropagation for training, the soft classification mode is enabled with $\bA^{\text{(soft)}}$ so as not to block the gradient flow. In other words, $\hat\bh=\bA^{\text{(soft)}}\bbeta$ represents each encoder output with a linear combination of all quantization bins. The process is summarized in Algorithm \ref{algo:quan}. Although this soft quantization process is differentiable and desirable during training, the discrepancy between $\bA^{\text{(soft)}}$ and $\bA^{\text{(hard)}}$ creates higher error during the test time, requiring a mechanism to reduce the discrepancy as in the following section. 

\subsubsection{Soft-to-hard quantization penalty}

Although the limit of $\bA^{\text{(soft)}}$ is $\bA^{\text{(hard)}}$ as $\alpha$ approaches $\infty$, the change of $\alpha$ should be gradual to allow gradient flows in the initial phase of training. We control the \textit{hardness} of $\bA^{\text{(soft)}}$ using the soft-to-hard quantization loss derived from \cite{new_paper_bloomberg}:
\begin{equation}\label{eq:softversushard}
    \calL_\text{Q}=\frac{1}{N}\sum_{n,k} \sqrt{\bA^{\text{(soft)}}_{nk}},
\end{equation}
whose minimum, 1, is achieved when $\bA^{\text{(soft)}}_{n:}$ is a one-hot vector for all $n$. Conversely, when $\bA^{\text{(soft)}}_{nk}=1/K$, the loss is maximum. Hence, by minimizing this soft-to-hard quantization penalty term, we can regularize the model to have \textit{harder} $\bA^{\text{(soft)}}$ values by updating $\alpha$ and the other model parameters accordingly. As a result, the test time quantization loss will be reasonably small when $\bA^{\text{(soft)}}$ is replaced by $\bA^{\text{(hard)}}$.

\subsubsection{Bitrate calculation and entropy control}
The bitrate is calculated as a product of the number of code values per second and the average bit depth for each code. The former is defined by the dimension of the code vector $N$ multiplied by the number of frames per second, $\frac{F}{T-o}$, where $T$, $o$, and $F$ are the input frame size, overlap size, and the original sampling rate, respectively. If we denote the average bit depths per sample by a function $g(\tilde{h}_n)$, the bitrate can be computed as in Eq. \eqref{eq:br},
\begin{equation}
\label{eq:br}
    \text{bitrate} = g(\tilde{h}_n)NF/(T-o). 
\end{equation}
When $F=16,000$, $T=512$, $o=32$, and $N=256$ after downsampling, for example, there are  about $8,533$ samples per second. If $g(\tilde{h}_n)=3$ bits, the bitrate is estimated as $25.6$ kbps. \zhenk{In contrast, the uncompressed bitrate is $256$ kbps with $N=T=512$, $o=0$, and $g(x_t)=16$ bits for each sample. }


\zhenk{We alter the entropy of $\bbeta$ to adjust the codec's bitrate, since the entropy serves as the lower bound of $g(\tilde{h}_n)$ based on Shannon's entropy theory. The entropy, $\mathcal{H}(\bbeta)$, is estimated with the sample distribution,
\begin{equation}\label{eq:ent1}
    \mathcal{H}(\bbeta) \approx -\sum_{k=1}^K p(\beta_k)\log_2 p(\beta_k),
\end{equation}
where $p(\beta_k)=\frac{1}{N}\sum_n\bA^{\text{(hard)}}_{nk}$ is the relative frequency of the $k$-th centroid being chosen during quantization.}
\zhenk{To navigate the model training towards the target bitrate,  $\mathcal{H}(\bbeta)$ defined in Eq. \eqref{eq:ent1} is included in the loss function as a regularizer: if the current  bitrate is higher than desired, the optimization process will increase the blending weight of the regularizer to strengthen the regularization effect, which consequently lowers the entropy, and vice versa. More details on the training target and hyperparameter setting are discussed in Sec. \ref{sec:hyperparameters}.}

\section{NWC-based Speech Coding Systems}
\label{sec:dsp}
\zhenk{By having NWC introduced in Sec. \ref{sec:neuralcodec} as the basic module, we propose two different extension mechanisms to improve the codec's performance in a wider range of bitrates, without increasing the model complexity significantly. 
First, in Sec. \ref{sec:cq}, we propose a neural network compatible LPC module where the trainable soft-to-hard quantization is applied to the LPC coefficients. With the LPC module followed by an NWC module, we achieve a win-win strategy that fuses the traditional DSP technique and the modern deep learning model \cite{zhen2020efficient}. In addition to integrating LPC, our neural codec conducts multistage residual coding \cite{zhen2019cascaded} by cascading residuals among multiple NWC modules (Sec. \ref{sec:cmrl}). The proposed CMRL system relays residual signals among the series of NWCs to scale up the coding performance at high bitrates.
}

\subsection{Trainable LPC Analyzer}
\label{sec:cq}
LPC has been widely used  to facilitate speech compression and sysnthesis, where \textit{source-filter} model ``explains out" the envelope of a speech spectrum, leaving a low-entropy residual signal \cite{ItakuraF1990lpc}. Similarly, LPC serves as a pre-processor in our system before its residual signal being compressed by NWC as we will see in Sec. \ref{sec:cmrl}. In this subsection, we redesign the LPC coefficient quantization process as a trainable module. We introduce collaborative quantization (CQ) to jointly optimize the LPC analyzer and NWCs as a residual coder. 



\subsubsection{Speech resonance modeling}
In the speech production process, the source 
as wide-band excitation signals go through the vocal tract tube. The shape-dependent resonances of the vocal tract  filter the excitations before it being transformed to speech signals \cite{titze1998principles}. In speech coding, the ``vocal tract response" is often modeled as an all-pole filter \cite{lpc}.
Having that, the $t$-th sample $x_t$ can be approximated by a autoregressive model using $M$ previous samples, 
\begin{equation}
    x_t = \sum\limits_{k=1}^{M}l_kx_{t-k} + e_t,
\end{equation}
where the estimation error $e_t$ represents the LPC residual, and $l_k$ denotes the filter coefficients.
Typically, ${l_k}$ can be efficiently estimated via Levinson-Durbin algorithm \cite{Levinson-Durbin}, and are to be quantized before LPC residual is calculated, i.e., $e_t$ encompasses the quantization error.
The LPC residual $e_t$ serves as input to the NWC module, which works as explained in Sec. \ref{sec:neuralcodec}, but on $\bolde$ rather than $\bx$. Hence, how LPC coefficients are quantized determines NWC's input, the LPC residual.

\subsubsection{Collaborative quantization}
The conventional LPC coefficient quantization process is standardized in  ITU-T G.722.2 (AMR-WB) \cite{ITUamrwb}: $2.4$k bits are assigned to represent the LPC coefficient per second though multistage vector quantization (MSVQ) \cite{gersho1983vector} in a classic LPC analyzer. Once again, we employ the soft-to-hard quantizer as illustrated in Sec. \ref{sec:neuralcodec} to make the quantization and bit allocation steps in the LPC analyzer trainable and communicatable with the neural codec. 





 \begin{figure}[t]
\centering
\subfigure[The trainable LPC analyzer where the linear spectral pairs are quantized by our trainable soft-to-hard quantizer (in the dotted box).]{\includegraphics[width=.8\columnwidth]{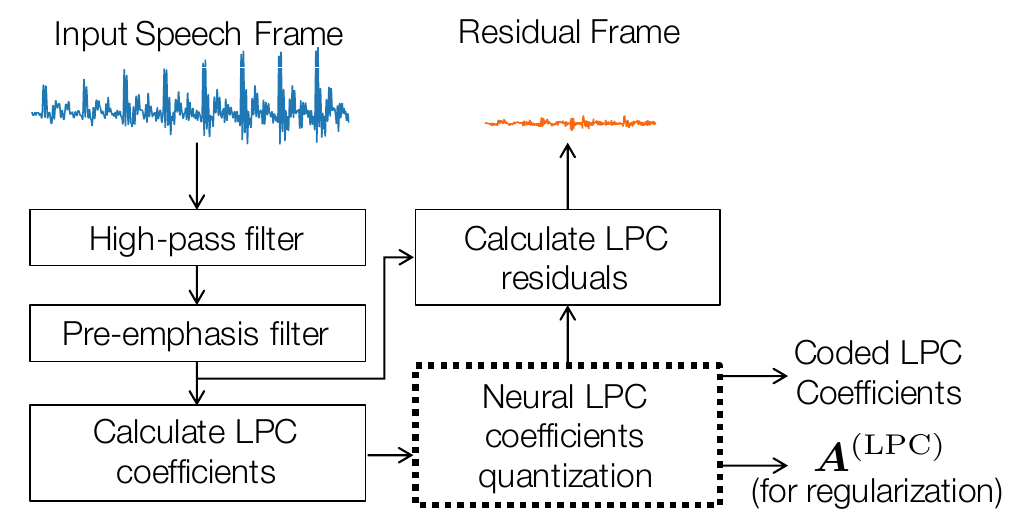}}
\label{fig:lpc_diagram}
\hspace{0in}

\subfigure[Cross-frame windowing]{\includegraphics[scale=0.45]{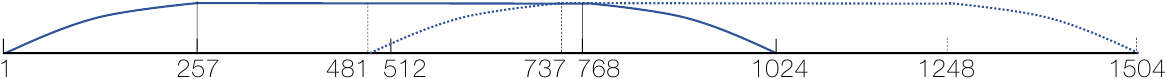}}
\subfigure[Sub-frame windowing]{\includegraphics[scale=0.45]{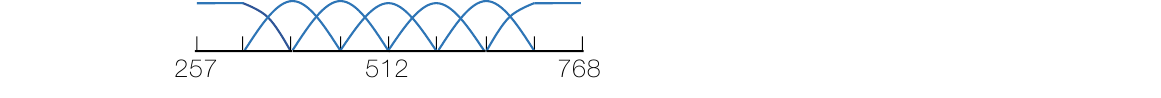}}
\subfigure[Synthesis windowing]{\includegraphics[scale=0.45]{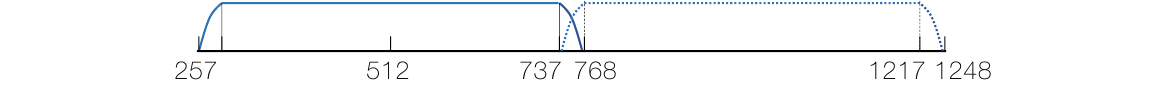}}
\caption{The signal flow chart for LPC analyzer (a) and windowing schemes in LPC (b)-(d).}
\label{fig:windowing}
\vspace{-0.05in}
\end{figure}

\zhenk{We compute the LPC coefficients as in \cite{amrwb}, first by applying high-pass filtering followed by pre-emphasizing (Fig. \ref{fig:windowing} (a)).} When calculating LPC coefficients, the window in Fig. \ref{fig:windowing} (b) is used. The window is symmetric with the left and right 25\% parts being tapered by a 512-point Hann window. After representing the 16 LPC coefficients in linear spectral pairs (LSP) \cite{soong1984line}, we quantize it using the soft-to-hard quantization scheme. Then, the sub-frame window in Fig. \ref{fig:windowing} (c) is applied to calculate LPC residual, which assures a more accurate residual calculation. The frame that covers samples [256:768], for instance, is decomposed into 7 sub-frames to calculate LPC residuals separately. 
Each 128-point Hann window in  Fig. \ref{fig:windowing} (c) is with 50\% overlap, except for the first and last window. They altogether form a constant overlap-add operation. Finally, after the synthesis using the reconstructed residual signal and corresponding LPC coefficients, the window  in Fig. \ref{fig:windowing} (d) tapers both ends of the synthesized signal, covering 512 samples with 32 overlapping samples between adjacent windows.

As an intuitive example, given the samples [1:1024] as the input, after the LPC analysis, neural residual coding, and LPC synthesis, samples [257:768] are decoded; the next input frame is [481:1504] (the dotted window in Fig. \ref{fig:windowing} (b)), whose decoded samples are within [737:1248]. The overlap-add operation is applied to the final decoded samples [737:768] (Fig. \ref{fig:windowing} (d)).

During this process, the calculated LPC coefficients are quantized using Algorithm \ref{algo:quan}, where the code vector is with 16 dimensions, i.e., $\bh\in\Real^{16}$. The number of kernels is set to be $K=2^8=256$. Note that the soft assignment matrix for the LPC quantization, $\bA^{(\text{LPC})}$, is also involved in the loss function to regularize the bitrate. 

We investigate the impact of the trainable LPC quantization in collaboration with the rest of the NWC modules in Sec. \ref{sec:eval}.

\subsection{Cross-Module Residual Learning (CMRL)}
\label{sec:cmrl}
To achieve scalable coding performance towards transparency at high bitrates, we propose cross-module residual learning (CMRL) to conduct bit allocation among multiple neural codecs in a cascaded manner. CMRL can be regarded as a natural extension of what is described in Sec. \ref{sec:cq}, where the LPC as a codec conducts the first round of coding by only modeling the spectral envelope. It leaves the residual signal for a subsequent NWC to be further compressed. With CMRL, we employ the concept of residual coding to cascade more NWCs. We also present a dual-phase training scheme to effectively train the CMRL model.

CMRL's scalability comes from its residual coding concept that enables a concatenation of multiple autoencoding modules. We define the residual signal recursively: $i$-th codec takes the residual of its predecessor as input, and the $i$-th reconstruction creates another residual for the next round, and so on. Hence, we have
\begin{equation}\label{eq:residual_rec}
    \hat\bx^{(i)}\leftarrow\calF^{(i)}(\bx^{(i)}), ~\bx^{(i)}\leftarrow\bx^{(i-1)}-\hat{\bx}^{(i-1)}, ~  \bx^{(1)}\leftarrow \bx,
\end{equation}
where $\hat\bx^{(i)}$ stands for the reconstruction of the $i$-th input using the $i$-th coding module $\calF^{(i)}(\cdot)$, while the input to the first codec is defined by the raw input frame $\bx$. If we expand the recursion, we arrive at the non-recursive definition of $\bx^{(i)}$,
\begin{align}
\label{eq:cmrl2}
    \bx^{(i)} = \bx - \sum\limits_{j=1}^{i-1}\hat{\bx}^{(j)},
\end{align}
which means the input to $i$-th model is the residual of the sum of all preceding $i-1$ codecs' decoded signals. 
It ensures the additivity of the entire system: adding more modules keeps improving the reconstruction quality. Hence, CMRL can scale up to high bitrates at the cost of increased model complexity.


CMRL is optimized in two phases. During Phase-I training, we sequentially train each codec from the first to the last one using a module-specific residual reconstruction goal, 
\begin{equation}\label{eq:cmrl_module_loss}
\calE(\bx^{(i)}||\hat{\bx}^{(i)}).     
\end{equation}
The purpose for Phase-I training is to get parameters for each codec properly initialized. Then, Phase-II finetunes all trainable parameters of the concatenated modules to minimize the global reconstruction loss,
\begin{equation}\label{eq:cmrl1}
\calE\left(\bx\bigg|\bigg| 
\sum_{i=1}^{N}\hat{\bx}^{(i)}
\right).
\end{equation}

\zhenk{The reconstruction loss measures the waveform discrepancy in both time and frequency domains. Quantization penalty and entropy control are introduced as regularizers. Sec. \ref{sec:hyperparameters} details the definition of the training target and ablation tests on how to find the optimal blending weights between the loss terms. }


 \begin{figure}[t]
\centering
\includegraphics[width=\columnwidth]{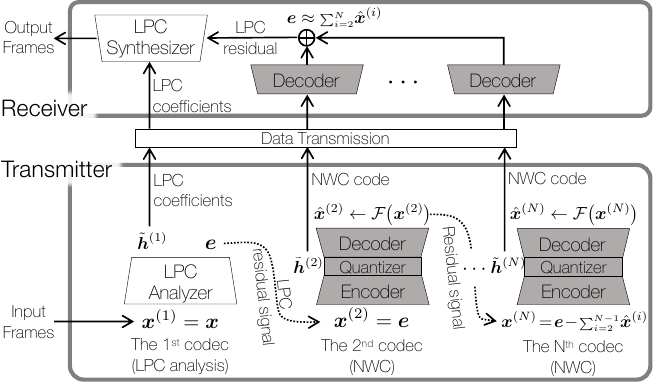}\hspace{0.0in}
\caption{The flow diagram of the test-time inference.}
\vspace{-0.1in}
 \label{fig:inference}
 \end{figure}

\subsection{Signal Flow during Inference}
Fig. \ref{fig:inference} shows the full system signal flow with $N$ sub-codecs, having an LPC module as the first one. On the transmitter side the LPC analyzer first processes the input frame $\bx$ of \zhenk{1024} samples and computes 16 coefficients, $\bh^{(1)}$, as well as \zhenk{512} residual samples $\bx^{(2)}$ \zhenk{at the center of the frame}. Then, the residual signal goes through the $N-1$ NWCs in sequentially. Note that the transmission process's primary job is to produce a quantized bitstring $\tilde{\bh}^{(i)}$ from LPC and each NWC.  To this end, NWC's decoder part must also run to compute the residual signal and relay it to the next NWC module. The bitstring is generated as a concatenation of all encoder outputs:  $\tilde{\bh}=\big[{\tilde{\bh}^{(1)}};{\tilde{\bh}^{(2)}}; \cdots; {\tilde{\bh}^{(N)}}\big]$. 
Once the bitstring is available on the receiver side, all NWC decoders run to reconstruct the LPC residual signal, i.e., $\hat\bx^{(2)}\approx\sum_{i=2}^N\calF_\text{dec}^{(i)}(\tilde{\bh}^{(i)})$. Then it is used as input of the LPC synthesizer, along with the LPC coefficients.

\section{Evaluation}
\label{sec:eval}
In this section, we examine the proposed neural speech coding model presented in Sec.\ref{sec:neuralcodec} and \ref{sec:dsp}. 
The evaluation critera include both objective measures such as PESQ \cite{pesq} and signal-to-noise ratio (SNR) and subjective scores from MUSHRA listening tests \cite{mushra}.
In addition, we conduct ablation analysis to provide a detailed comparison between various loss terms and bit allocation schemes. Finally, we report the system delay and execution time under four hardware specifications.

\subsection{Data Processing}
The training dataset is created from 300 speakers randomly selected from the TIMIT  corpus \cite{timit} with no gender preference. Each speaker contributes 10 utterances totaling 2.6 hour-long training set, which is a reasonable size due to our compact design.
The same scheme is adopted when creating the validation dataset and test dataset with 50 speakers, respectively. \zhenk{All three datasets are mutually exclusive with the sample rate of $16$ kHz.}
All neural codecs in this work are trained and tested via the same set of data for a fair comparison. 
We normalize each utterance to have a unit variance, then divided by the global maximum amplitude, before being framed into segments with the size of 512 samples. On the receiver side, we conduct overlap-and-add after the synthesis of the frames, where a 32-sample Hann window is applied to the overlapping region of the same size.

With the LPC codec, we apply high-pass filtering defined in the z-space, $\mathcal{G}_\text{hp}(z)\!=\frac{0.989502 - 1.979004z^{-1} 0.989502z^{-2}}{1 - 1.978882^{-1} 0.979126^{-2}}$, to the normalized waveform. A pre-emphasis filter, $\mathcal{G}_\text{preemp}(z)\!=1\!-\!0.68z^{-1}$, follows to boost the high frequencies.

\subsection{Training Targets and Hyperparameters} \label{sec:hyperparameters}
\begin{figure}[t]
\centering
\includegraphics[width=\columnwidth]{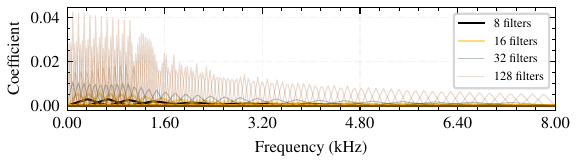}
\caption{The coarse-to-fine filter bank analysis in the mel scale.}
\vspace{-0.1in}
 \label{fig:mel-1}
 \end{figure}

The loss function is defined as 
\begin{equation}
\begin{split}
\mathcal{L} &= \lambda_{\text{MSE}}\sum_{t=1}^T(x_t-\hat{x}_t)^2 + \lambda_{\text{mel}} \sum_{b=1}^4\sum_{f=1}^{F_b}\Big(y_f^{(b)}-\hat y_f^{(b)}\Big)^2\\
& +\lambda_{\text{Q}} \calL_\text{Q} + \zhenk{\lambda_{\text{ent}}} \mathcal{H}(\bbeta)\\
\end{split}
\label{eq:loss}
\end{equation}
where the first term measures the mean squared error (MSE) between the raw waveform samples and their reconstruction. Ideally, if the model complexity and the bitrate is sufficiently large, an accurate reconstruction is feasible by using MSE as the only loss function. 
Otherwise, the result is usually sub-optimal due to the lack of bits: coupled with the MSE loss, the decoded signals tend to contain broadband artifact.
The second term supplements the MSE loss and helps suppress this kind of artifact. To this end, we follow the common steps to conduct mel-scaled filter bank analysis, which results in a mel spectrum $\by$ that has a higher resolution in the low frequencies than in the high frequencies. 
The filter bank size defines the granularity level of the comparison. Following \cite{new_paper_bloomberg}, we conduct a coarse-to-fine filter bank analysis by setting four filter bank sizes, $F_1=8, F_2=16, F_3=32, F_4=128$ as shown in Fig. \ref{fig:mel-1}, which result in four kinds of resolutions for mel spectra $\by^{(b)}$ indexed by $b\in\{1,2,3,4\}$.





All models are trained on Adam optimizer with default learning rate adaptation rates \cite{adam}. The batch size is fixed with $128$ frames . The initial learning rate is $2\times 10^{-3}$ for the first neural codec.
With CMRL, the learning rate for the successive neural codecs is $2\times 10^{-4}$. Finetuning of all those models is with a smaller learning rate $2\times 10^{-5}$. All models are sufficiently trained until the validation loss converges after being exposed to about  $5\times 10^{5}$ batches. These hyperparameters were chosen based on validation.

The blending weights in the loss function in Eq. \eqref{eq:loss} are also selected based on the validation performance. Empirically, the ratio between the time-domain loss and mel-scaled frequency loss affects the trade-off between the SNR and perceptual quality of decoded signals. If the time-domain loss dominates the optimization process, the model compresses each sub-band with an equal effort. In that case, the artifact will be audible unless the SNR reaches a rather high level (over $30$ dB) which entails a high bitrate and model complexity. On the other hand, if only the mel-scaled frequency loss is in place, the reconstruction quality in the high frequency will degrade.  The impact of these blending weights for these two loss terms is detailed in Sec. \ref{sec:ablation} via an ablation analysis. 

The weights for the quantization regularizer $\lambda_\text{Q}$ and entropy regularizer $\lambda_\text{ent}$ are initially set to be $0.5$ and $0.0$, respectively. \zhenk{As for $\lambda_\text{ent}$, we alter it after every epoch by $0.015$: if the current model's bitrate is higher than the target bitrate, $\lambda_\text{ent}$ increases to penalize the model's entropy more; otherwise, $\lambda_\text{ent}$ decreases to boost the entropy.} Note that we omit the module index $i$ in Eq. \eqref{eq:loss}, so the meaning of $\hat x_t$ depends on the context: either the module-specific reconstruction as in Eq. \eqref{eq:cmrl_module_loss} or the sum of all recovered residual signals for Phase-II finetuning as in Eq. \eqref{eq:cmrl1}. Similarly, $\calL_\text{Q}$ and $\calH_{\bbeta}$ can encompass all modules' quantization and entropy losses including LPC's for Phase-II. We delay the introduction of the quantization and entropy loss until the fifth epoch. 



\subsection{Bitrate Modes and Competing Models}
We consider three bitrates, $12$, $20$, and $32$ kbps, to validate models' performance in a range of use cases. We evaluate following different versions of neural speech coding systems: 
\begin{itemize}
    \item Model-I: The NWC baseline (Sec. \ref{sec:neuralcodec}).
    \item  Model-II: Another baseline that combines the legacy LPC and an NWC module for residual coding.
    \item  Model-III: A trainable LPC quantization module followed by an NWC and finetuning (Sec. \ref{sec:cq});
    \item  Model-IV: Similar to Model-III but with two NWC modules: the full-capacity CMRL implementation (Sec.\ref{sec:cmrl}). It is tested cover the high bitrate case, $32$ kbps.
\end{itemize}
Regarding the standard codecs, AMR-WB \cite{amrwb} and Opus \cite{opus} are considered for comparison. AMR-WB, as an ITU standard speech codec, operates in nine different modes covering a bitrate range from 6.6 kbps to 23.85 kbps, providing excellent speech quality with a bitrate as low as 12.65 kbps in wideband mode. As a more recent codec, Opus shows the state-of-the-art performance in most bitrates up to 510 kbps for stereo audio coding, except for the very low bitrate range.

We first compare all models with respect to the objective measures, while being  aware  that they are not consistent with the subjective quality. Hence, we also evaluate these codecs in two rounds of MUSHRA subjective listening tests: the neural codecs are compared in the first round, whose winner is compared with other standard codecs in the second round.



\begin{table*}[t]
\centering
\caption{Objective measurements for neural codec comparison under three bitrate cases.}
\resizebox{\textwidth}{!}{
\begin{tabular}{c|cccccc|cccccc}
\toprule
Bitrate  &\multicolumn{6}{c|}{SNR (dB)} & \multicolumn{6}{c}{PESQ-WB} \\
(kbps)& Model-I& Model-II& Model-III& Model-IV & AMR-WB & Opus & Model-I& Model-II& Model-III& Model-IV & AMR-WB & Opus \\
\midrule
$\sim$12&  \textbf{12.37}    &  10.69    &  10.85    &  --  &   11.60       &    9.63    &  3.67     &  3.45 &  3.60    & -- &  3.92    & \textbf{3.93} \\
$\sim$20&   \textbf{16.87}   &  10.73    &  13.65    &  --  &   13.14   &  9.46    & \textbf{4.37}     &  3.95 &  4.01    & -- &   4.18   & \textbf{4.37} \\
$\sim$32&   \textbf{20.24}   &  11.84    &     14.46      &   17.11   &    --     & 17.66     &  \textbf{4.42}    &  4.15   &  4.18  &  4.35  &   --   & 4.38 \\
\bottomrule
\end{tabular}}
\label{tab:ob_comp}
\end{table*}
 
\subsection{Objective Measurements}

\subsubsection{The compact NWC module and its performance}\label{sec:nwc_analysis}
\begin{figure}[t]
    \centering
    \subfigure[The validation SNR curve during training]{\includegraphics[width=.5\textwidth]{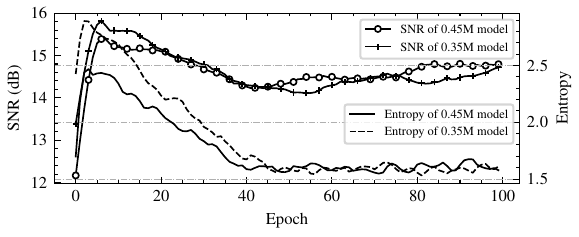}}
    \hspace{-3pt}
    \subfigure[The validation PESQ curve during training]{\includegraphics[width=.5\textwidth]{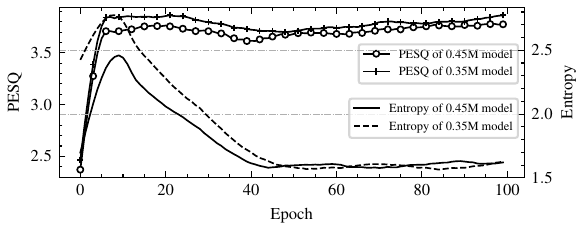}}
    \caption{Speech reconstruction performance stays almost the same when the model size decreases from 0.45 to 0.35 million parameters with the help from the structural modification.}
    \hspace{-3pt}
    \label{fig:35-45-comp}
\end{figure}

Compared to our previous models in \cite{zhen2019cascaded}\cite{zhen2020efficient} that use $0.45$ million parameters, the newly proposed NWC in this work only has $0.35$ million parameters. It is also a significant reduction from the other compact neural waveform codec \cite{new_paper_bloomberg} with $1.6$ million parameters. As introduced in Sec. \ref{sec:neuralcodec} the model size reduction is achieved via the GLU \cite{tan2019gated} and depthwise separable convolution for upsampling \cite{chollet2017xception}.  In our first experiment, we show that the objective measures stay the same. Fig. \ref{fig:35-45-comp} compares the NWC modules before and after the structural modification proposed in Sec. \ref{sec:neuralcodec} in terms of (a) signal-to-noise ratio (SNR) and (b) PESQ-WB \cite{pesq}. We can see that the newly proposed model with $0.35$M parameters is comparable to the larger model. Therefore, it justifies its use as the basic module in a range of models from Model-I to IV.


\begin{figure}[t]
\centering
 \label{fig:cmrl-gt-curves-snr}
\subfigure[Scalability with respect to SNR]{\includegraphics[width=0.49\textwidth]{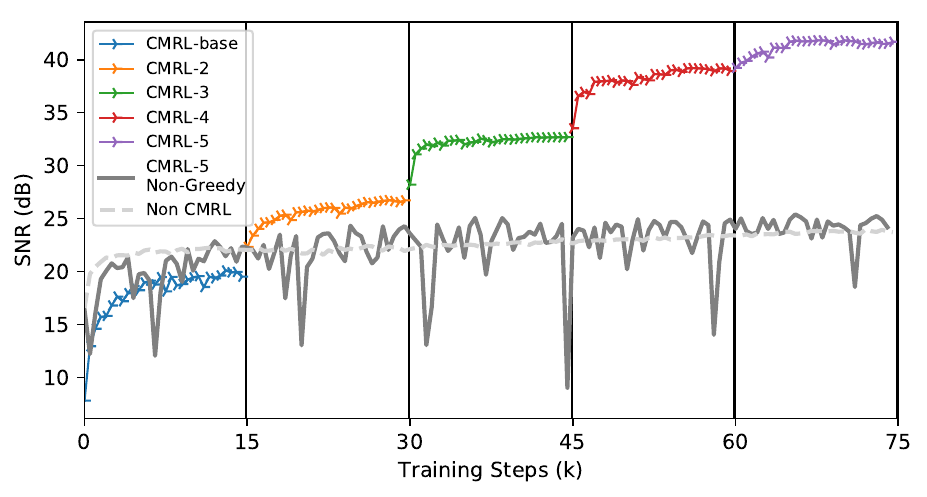}}
\label{fig:cmrl-gt-curves-pesq}
\subfigure[Scalability with respect to PESQ ]{\includegraphics[width=0.49\textwidth]{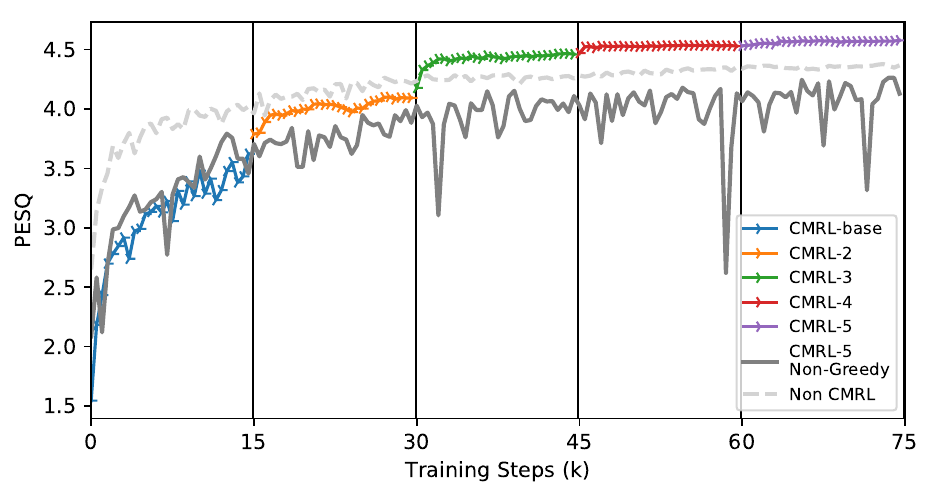}}
\hspace{0.0in}
\caption{In CMRL, performance leaps when the new neural codec is added for residual cascading. \zhenk{In these CMRL models LPC is not used.} }
 \label{fig:cmrl-gt-curves}
 \end{figure}

\subsubsection{The impact of CMRL's residual coding}\label{sec:cmrl_analysis}

To validate the merit of CMRL's residual coding concept, we scale up the CMRL model by incrementally adding more NWC modules up to five. In Fig. \ref{fig:cmrl-gt-curves}, both SNR and PESQ values keep increasing when CMRL keeps adding a new NWC module. There are two noticeable points in these graphs. First, the greedy module-wise pretraining is important for the performance: whenever a new model is added, it is pretrained to minimize the module specific loss Eq. \eqref{eq:cmrl_module_loss} first (Phase-I), then the global loss Eq. \eqref{eq:cmrl1}, subsequently (Phase-II). A model that does not perform Phase-II (thick gray line) stagnates no matter how many NWCs are added. Second, we also train a very large NWC model with the same amount of parameters as CMRL with five NWCs combined (grey dash). It turns out the equally large model fails to scale up due to its single integrated architecture. While we eventually decide to use only up to two NWCs for speech coding for our highest bitrate case, $32$ kbps, \zhenk{we may keep adding NWCs to CMRL to meet the case of higher bitrates for non-speech audio coding.}



\subsubsection{Overall objective comparison of all competing models}
TABLE \ref{tab:ob_comp} reports SNR and PESQ-WB from all competing systems. AMR-WB in the low-range bitrate setting operates at 12.65 kbps and 23.05 kbps for the mid-range. 
\zhenk{Among neural speech coding systems, Model-I,  as a single autoencoder model, outperforms others in all three bitrate setups in terms of SNR and PESQ-WB. It is also comparable to AMR-WB and Opus, except for the low bitrate case where Opus achieves the highest PESQ score.
One explanation is that Model-I is highly optimized for the objective loss during training, although it does not necessarily mean that the higher objective score leads to a better subjective quality as presented in Sec. \ref{sec:mos}.}
It is also observed that with CQ, Model-III gains slightly higher SNR and PESQ scores compared to Model-II, which uses the legacy LPC. Finally, the performance scales up significantly when Model-IV starts to employ two NWCs on top of LPC, which is our proposed full neural speech coding setup. Aside from objective measure comparison, to further evaluate the quality of proposed codec, we discuss the subjective test in the next section. 


 \begin{figure}[t]
\centering
\vspace{-.1in}
\subfigure[Neural waveform codecs comparison.]{\includegraphics[width=\columnwidth]{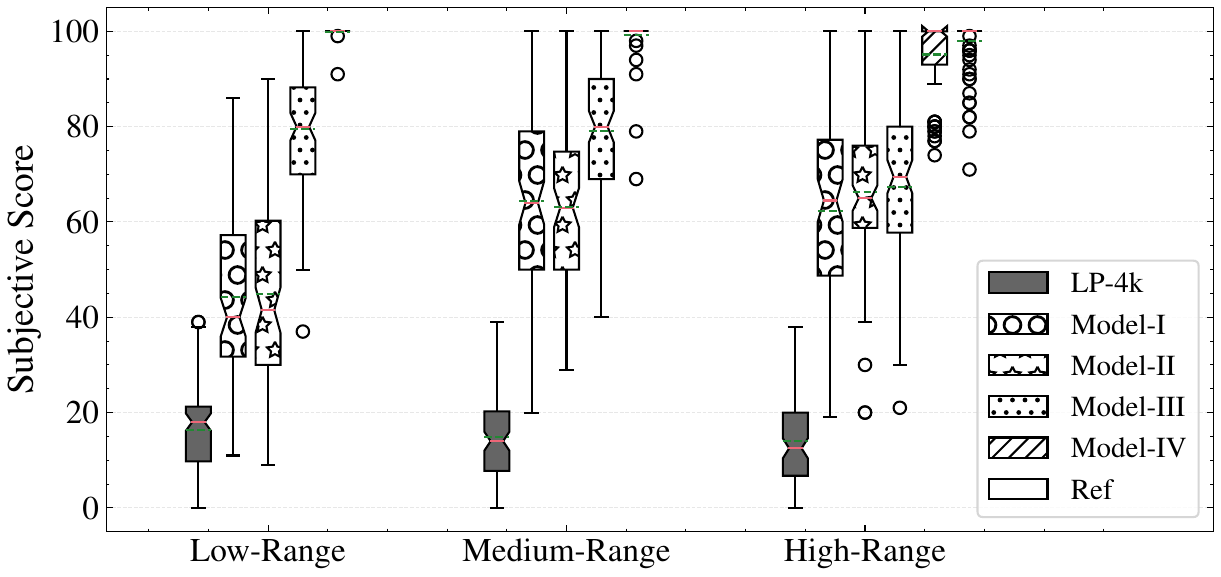}\hspace{0.0in}}
\vspace{-.1in}
\subfigure[Comparison between proposed methods and standard codecs.]{\includegraphics[width=\columnwidth]{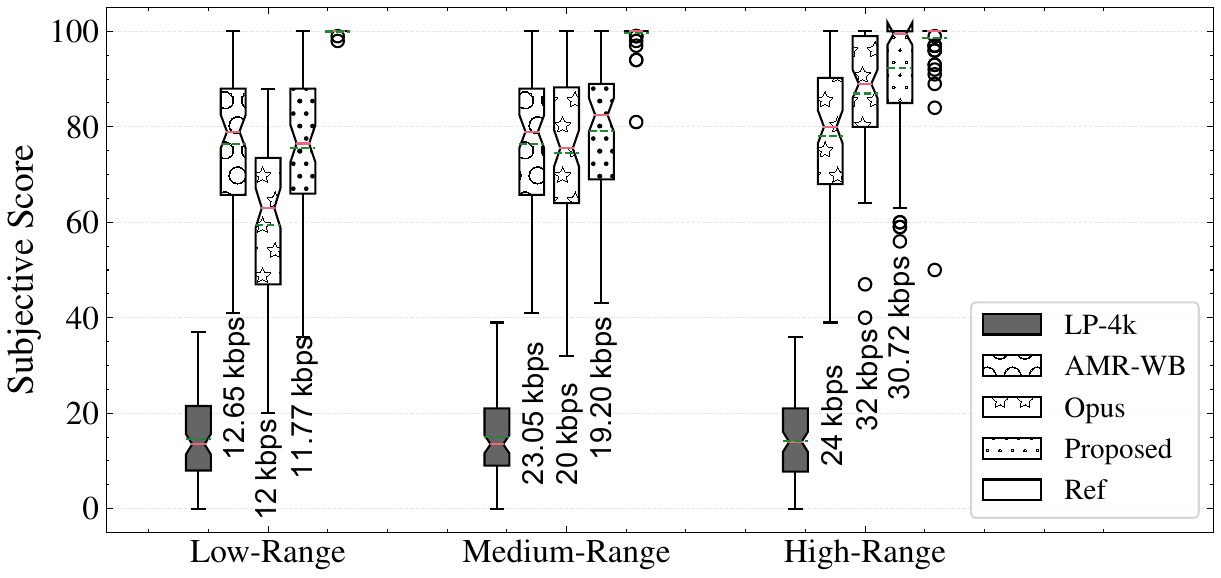}\hspace{0.0in}}
 \caption{MUSHRA subjective listening test results.}
 \label{fig:box}
 \vspace{-.1in}
 \end{figure}


\subsection{Subjective Test}
\label{sec:mos}
We conduct two rounds of MUSHRA tests: (a) to select the best one out of the proposed models (from Model-I to IV) (b) to compare it with the standard codecs, i.e., AMR-WB and Opus. Each round covers three different bitrate ranges, totaling six MUSHRA sessions. A session consists of ten trials, for which ten gender-balanced test signals are randomly selected. Each trial has one low-pass filtered signal serving as the anchor (with a cutoff frequency at 4kHz), the hidden reference, as well as signals decoded from competing systems. We recruit ten participants who are audio experts with prior experiences in speech/audio quality evaluation. The subjective scores are rendered in Fig. \ref{fig:box} as boxplots.  
Each box ranges from the $25$ to  $75$ percentile  with a $95\%$ confidence interval.
The mean and median are presented as the green dotted line and pink hard line, respectively. Outliers are represented in circles.

\begin{figure*}
\begin{minipage}{.438\textwidth}
\vspace{0.05in}
{\includegraphics[width=\linewidth]{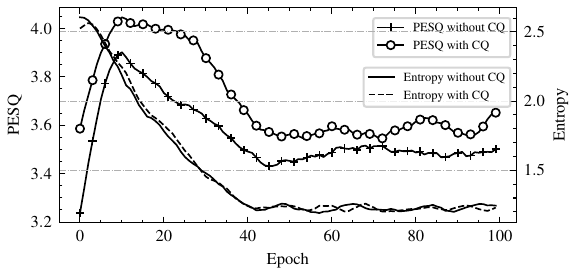}}
\vspace{-0.2in}
    \caption{The ablation analysis on CQ.}
    \label{fig:cq}
\end{minipage}    
\begin{minipage}{.56\textwidth}
\includegraphics[width=\linewidth]{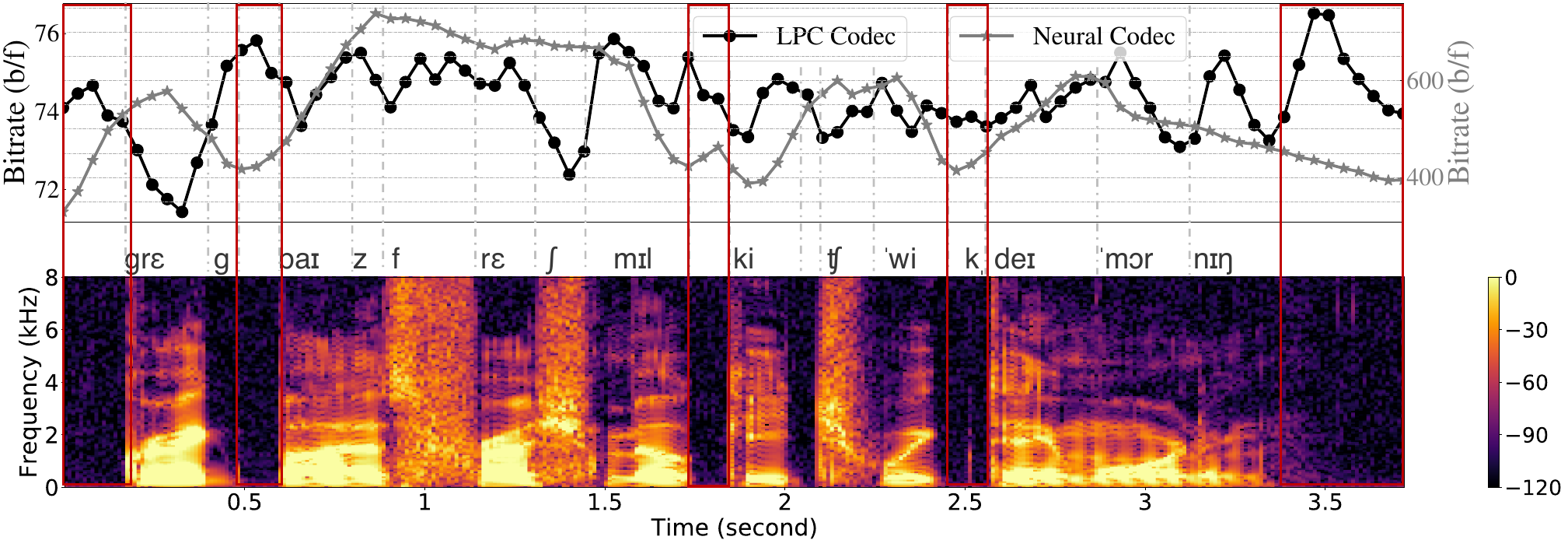}\hspace{0.0in}
\caption{The frame-wise bit allocation analysis. }
\vspace{-0.1in}
 \label{fig:wide-spec-rate}
\end{minipage}
\end{figure*}

\subsubsection{Comparison among the proposed neural codecs}
In Fig. \ref{fig:box} (a) we see that Model-III's produces decoding results that are much more preferred than both Model-I and Model-II, which are a pure end-to-end model and with the non-trainable legacy LPC module, respectively. The advantage is more noticeable in lower bitrates. It is contradictory to the objective scores reported in TABLE \ref{tab:ob_comp} where Model-I often achieves the highest scores. 
\zhenk{Compared to the deterministic quantization component in the legacy LPC in Model-II, LPC module and NWC in Model-III are jointly trained for a frame-wise independent bit allocation, so as to maximize the coding efficiency.} We also note that Model-III's performance stagnates in the high bitrate experiments, suggesting its poor scalability. To this end, for the high bitrate experiment, we additionally test Model-IV with two NWC residual coding modules instead of just one. \zhenk{Model-IV outperformed both Model-II and Model-III by a large margin, showcasing a near-transparent quality. }

\subsubsection{Comparison with standardized codecs}
Fig. \ref{fig:box} (b) shows that, our Model-II is on par with AMR-WB for the low-range bitrate case, while outperforming Opus which tends to lose high frequency components. In the medium-range, Model-II at $19.2$ kbps is comparable to Opus at $20.0$ kbps and AMR-WB at $23.05$ kbps. In the high bitrate range, our Model-IV outperforms Opus that operates  $32$ and $24$ kbps, while AMR-WB is omitted as it does not support those high bitrates. 

\zhenk{All MUSHRA sessions are available online, along with demo samples and source codes\footnote{\href{https://saige.sice.indiana.edu/research-projects/neural-audio-coding}{https://saige.sice.indiana.edu/research-projects/neural-audio-coding}}}.


\begin{table}[t]
\caption{Ablation analysis on blending weights.} \label{tab:mel_ablation}
\centering
\begin{subtable}{(a) \zhenk{Model-I (a neural codec only)}}\\
\centering
\begin{tabular}{c|cc}
\toprule
Blending Ratio (MSE : mel) & Decoded SNR (dB) & Decoded PESQ  \\
\midrule
$1:0$ &           \textbf{18.12}    & 3.67           \\
$0:1$ &           0.16    & 4.23           \\
$1:1$ &           6.23    & 4.31           \\
$10:1$ &          16.88        & \textbf{4.37}           \\
 \bottomrule
\end{tabular}
\end{subtable}
\begin{subtable}{(b) \zhenk{Model-III (a neural codec with a collaboratively trained LPC)}}
\centering
\resizebox{\columnwidth}{!}{
\begin{tabular}{c|ccc}
\toprule
Blending Ratio & Residual SNR  & Decoded SNR & Decoded PESQ \\
(MES : mel) & (dB) & (dB) &\\
\midrule
$1:0$ & \textbf{9.73}                   & 14.25               & 3.84           \\
$0:1$ & 1.79                    & 17.23               & 4.02           \\
$1:1$ & 7.11                    & \textbf{17.82}                & \textbf{4.08}           \\
$10:1$ & 8.26                    & 17.55               & 4.01           \\
 \bottomrule
\end{tabular}}
\end{subtable}
\label{tab:ablation_loss}
\end{table}

\subsection{Ablation Analysis}
\label{sec:ablation}
In this section, we perform some ablation analyses to justify our choices that led to CQ and CMRL's superior subjective test results. We investigate how different blending ratios between loss terms can alter the performance. We will also explore the optimal bit allocation strategy among coding modules.

\subsubsection{Blending weights for the loss terms}\label{sec:loss_abl}
Out of the two reconstruction loss terms, MSE serves as the main loss for the end-to-end NWC system, while the mel-scaled loss prioritizes certain frequency bands over the others. \zhenk{We perform ablation analysis on four blending ratio settings to analyze their effect on decoded speech's objective quality. We consider both system configurations: one with only the neural waveform codec (TABLE \ref{tab:ablation_loss} (a)) and the other one with both the neural waveform codec and collaboratively trained LPC module (TABLE \ref{tab:ablation_loss} (b). 
For TABLE \ref{tab:ablation_loss} (a), the target entropy for each sample in the neural codec is of $2.5$-bit, corresponding to a bitrate of $\frac{512-32}{16000*1024}*256*2.5\approx21.3$ kbps. 
The SNR reaches the highest when there is only the MSE term, while the PESQ score becomes the lowest. By only keeping the mel-scaled loss term, the PESQ score is decent (4.23), yet with a poor waveform reconstruction as suggested by the SNR value (0.16 dB). 
For TABLE \ref{tab:ablation_loss} (b), the target entropy for each LPC coefficient is of $5$-bit or $2.6$ kbps, and $3.5$-bit for each LPC residual sample or $29.9$ kbps. Similarly, even with the input of the neural codec being the LPC residuals, MSE alone yields the highest SNR for the reconstruction of the LPC residual, but it does not benefit the final synthesized signal even in terms of SNR. Note that we choose 128 quantization centroids for the high bitrate case, which is different from that of Model-III in TABLE \ref{tab:ob_comp} where 32 quantization centroids are employed.}
For consistency's sake, we choose the blending ratio of $10:1$, which shows reasonably well numbers in all proposed models.


\subsubsection{CQ's impact on the speech quality}\label{sec:cq_abl}

We compare the PESQ values of the decoded signals from Model II and III. Since Model-III shares the same architecture with Model-II except for the CQ training strategy, the comparison is to verify that CQ can effectively allocate bits to the LPC and NWC modules. Fig. \ref{fig:cq} shows that the total entropy of the two models are under the control regardless of the use of CQ mechanism. However, we can see that Model-III with CQ achieves higher PESQ during and after the control of the entropy, showcasing that the CQ approach benefits the codec's performance.

\subsubsection{Bit allocation between the LPC and NWC modules}\label{sec:bit_alloc_lpc_nwc}

Since the proposed CQ method is capable of assigning different bits to the LPC and NWC modules dynamically, i.e., in a frame-by-frame manner, we analyze its impact in more detail. In the mid-range bitrate setting, Fig. \ref{fig:wide-spec-rate} shows the amount of bits assigned to both modules per frame (b/f). First of all, we observe that the dynamic bit allocation scheme indeed adjusts the LPC and NWC bitrates over time. 
\zhenk{Because of the CQ-enabled dynamic bit allocation, our system is able to compress silent frames more efficiently: by allocating just a few more bits to LPC, it saves a lot more bits from the NWC module for residual coding, as shown in the crimson-colored boxed areas in Fig. \ref{fig:wide-spec-rate}.}
However, it still requires a significant amount of bits to even represent those near-silent frames, which can be further optimized by voice activity detection. Finally, it appears that NWC is less efficient for fricatives (e.g., \textit{\textipa{f}} and \textit{\textipa{S}}) and affricates (e.g., \textit{\textipa{tS}}). TABLE \ref{tab:bit_allocation} shows the overall bit allocation among different modules. In the low bitrate case, it is worth noting that CQ uses 58 b/f or $1.93$ kbps, differently from AMR-WB's standard, $2.4$ kbps.

\subsubsection{Bit allocation between the two NWC modules in Model-IV} \label{sec:bit_alloc_nwc}
\begin{figure}[t]
\centering
\includegraphics[width=0.45\textwidth]{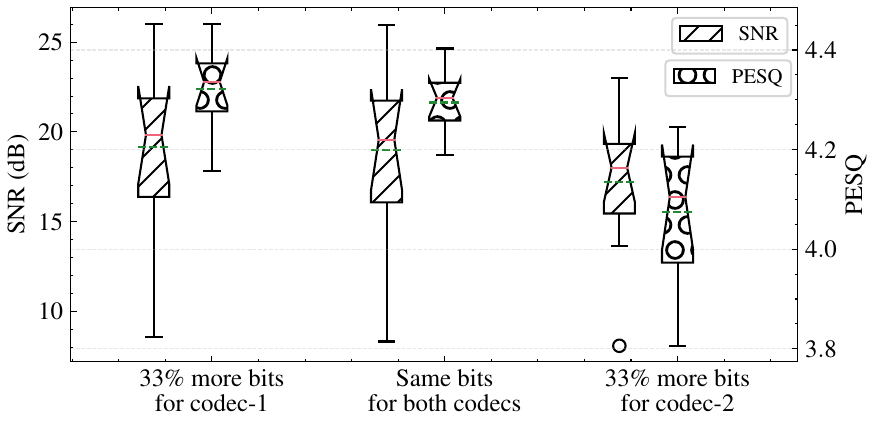}\hspace{0.0in}
 \caption{Ablation analysis on bit allocation schemes between codec-1 and codec-2 in Model-IV at $32$ kbps.}
 \label{fig:bit_allocation}
 \end{figure}

To find the optimal bit allocation between two NWC modules, we first conduct an ablation analysis on $3$ different bit allocation choices. In Fig. \ref{fig:bit_allocation}, both the SNR and PESQ scores degrade when the second NWC uses 33.3\% more bits than the first one. Among these $3$ choices, the highest PESQ score is obtained when the first NWC module uses 33.3\% more bits. In practice, the bit allocation is automatically determined during the optimization process. In TABLE \ref{tab:bit_allocation}, for example, the bit ratio between two NWC modules of Model-IV in the high bitrate case is about $486:384\approx55.9:44.1$, in accord with the observation from the ablation analysis that the first module should use more bits.


\begin{table}[t]
\centering
\caption{Bit allocation among coding components.}
\begin{tabular}{c|ccc}
\toprule
Bitrate Modes & LPC Coefficients  & LPC Residual& Total \\
(kbps) & (bits / frame) & (bits / frame)& (bits / frame)\\
\midrule
$\sim 11.77$ & 58                    & 295               & 353        \\
$\sim 19.20$ &  74                    & 502               & 576         \\
$\sim 30.72$ &  74                    & 486+384           & 944        \\ 
\bottomrule
\end{tabular}
\label{tab:bit_allocation}
\end{table}

\subsection{Complexity and Delay}\label{sec:complexity}
The proposed NWC model is with $0.35$ million parameters, a half of which is for the decoder. Hence, in $32$ kbps with two NWC modules for residual coding, the model size totals 0.7M parameters, with the decoder size of $0.35$M. Even though our decoder is still not as compact as those in traditional codecs, it is $100\times$ smaller than a WaveNet decoder.

Aside from the model size, we investigate the codec's delay and the execution induced latency. The codec will have algorithmic delay if it relies on future samples to predict the current sample. The processing time during the encoding and decoding processes also adds up to the runtime overhead.

\subsubsection{Algorithmic delay}
The delay of our system is defined by the frame size: the first sample of a frame can be processed only after the entire frame is buffered: \zhenk{$1024/16000=64$ms}. 
Causal convolution can minimize such delay at the expense of the reduced speech quality, because it only uses past samples.
 
 \begin{table}[t]
\centering
\caption{Execution time ratios during model inference (\%). }
\label{tab:exec_time}
\resizebox{\columnwidth}{!}{
\begin{tabular}{ c|cccc}
 \toprule
Hardware & 0.45M& 0.35M& 0.45M$\times$2 & 0.35M$\times$2\\
 \midrule
$1\times$ Tesla V100 & 12.49& 13.38& 20.69& 21.12\\
$1\times$ Tesla K80 & 24.45& 22.53& 39.42& 38.82\\
\midrule
$8\times$ CPU cores & 20.76& 18.91& 35.17& 33.80\\
$1\times$ CPU core & 46.88& 42.44& 87.38& 80.21\\
\bottomrule
\end{tabular}}
\vspace{0.15in}
\end{table}

\subsubsection{Analysis of the execution time} 
The execution time is another important factor to be considered for real-time communications. The bottom line is that the execution of the encoding and decoding processes is expected to be within the duration of the hop length so as not to lead to execution induced latency. For example, WaveNet codec \cite{KleijnW2018wavenet} minimizes the system delay using causal convolution, but its processing time, though not reported, can be rather high as it is an autoregressive model with over $20$ million parameters. TABLE \ref{tab:exec_time} lists the execution time ratio of our models. The ratio (in percentage) is defined as the execution time to encode and decode the test signals divided by the duration of those signals. Meanwhile, Kankanahalli's model requires $4.78$ms to encode and decode a hop length of $30$ms on an NVIDIA\textsuperscript{\textregistered}  GeForce\textsuperscript{\textregistered} GTX 1080 Ti GPU, and $21.42$ms on  an Intel\textsuperscript{\textregistered} Core\textsuperscript{TM} i7-4970K CPU (3.8GHz), which amount to $15.93\%$ and $71.40\%$ of the execution time ratio, respectively  \cite{new_paper_bloomberg}. Our small-sized models (0.45M and 0.35M) on both CPU (Intel\textsuperscript{\textregistered} Xeon\textsuperscript{\textregistered} Processor E5-2670 V3 2.3GHz) and GPU run faster than Kankanahali's, while the direct comparison is not fair due to the different computing environment. 
The CMRL models with two NWC modules require more execution time. 
Note that all our models compared in this test achieved the real-time processing goal as their ratios are under 100\%. 
\zhenk{The proposed NWC with 0.35 million parameters runs faster on CPUs than its predecessor \cite{zhen2019cascaded}\cite{zhen2020efficient} with 0.45 million parameters, although the comparison is not consistent on GPUs, due to TensorFlow's optimization effects at runtime.}



\zhenk{\subsubsection{Implementation notes and limitations}\label{sec:limitations}
While the proposed model noticeably reduces the decoding time and memory footprint compared to our previous models and the WaveNet decoder, it may not be ready to directly meet real-world requirements. 
Concretely, our neural codec can be implemented in a more hardware-friendly fashion, so as to allow the processor to handle multiple other tasks.
One promising direction is to quantize the neural network. For example, instead of using the single-precision (32 bit) floating points to represent model weights, we can represent each weight by one of 255 values (8-bit quantization), which enables much simpler arithmetic operations during inference. Furthermore, depending on the model architecture, pruning away less important weights is a sensible method to compress a network. Investigating these network compression methods for our codecs is beyond the scope of this paper, but it has been observed that quantization and pruning bring little to no degradation to neural networks for speech recognition \cite{HanS2017fpga_asr}. The recently proposed PercepNet architecture  also showed  high-quality, real-time speech enhancement is possible using less than $5\%$ of a CPU core for speech enhancement \cite{valin2020perceptually}. \zhenksep{The current system is not causal with an algorithmic delay of 64ms, which necessitates the employment of causal convolutions for real-time applications.}
}



\section{Concluding Remarks}
\label{sec:conclusion}
Recent neural waveform codecs have outperformed the conventional codecs in terms of coding efficiency and speech quality, at the expense of model complexity.
We proposed a scalable and lightweight neural acoustic processing unit for waveform coding. Our smallest model contains only $0.35$ million parameters whose decoder with $0.12$ million parameters is more than $160\times$ smaller than the WaveNet based codec. \zhenk{Having a compact design as a neural network, by incorporating a trainable LPC analyzer and residual cascading, 
our models reconstruct clean English speech samples with the quality on par with or superior to that from standardized codecs.} 
\zhenk{Admittedly, these standardized codecs are more computationally efficient and have reasonably well performed from narrow to full band scenarios already. It is still highly desired if these standalone DSP components can be reformulated into a lightweight but end-to-end trainable format for a full neural speech processing pipeline.}
\zhenk{To that end, the proposed system serves a candidate as it operates in a frame-wise manner with the processing time less than the frame length, even with a less optimized python-based implementations.}

\zhenk{The proposed system is still computationally heavier than traditional speech codecs. Therefore, running the model in embedded systems with limited computational resources may require further model compression, such as parameter quantization and pruning.
Although the neural waveform codec is generic and not contingent upon language specific priors, its generalizability to different languages and noisy and reverberant acoustic environments is not guaranteed, especially if the model sizes are relatively small.}

\zhenk{We open-sourced the project. The source code and sound examples can be found at: \href{https://saige.sice.indiana.edu/research-projects/neural-audio-coding}{https://saige.sice.indiana.edu/research-projects/neural-audio-coding}}.


%
\IEEEpeerreviewmaketitle

\ifCLASSOPTIONcaptionsoff
  \newpage
\fi

\bibliographystyle{IEEEtran}
\bibliography{mjkim}
\end{document}